\documentclass[12pt,preprint]{aastex}
\usepackage{natbib}

\shorttitle{Radio Photosphere and Mass-Loss Envelope of VY CMa}
\shortauthors{Lipscy, Jura, \& Reid}

\begin{document}
\title{Radio Photosphere and Mass-Loss Envelope of VY CMa}
\author{S.~J. Lipscy\altaffilmark{1,2}, M. Jura\altaffilmark{1},
M.~J. Reid\altaffilmark{3}}
\email{slipscy@ball.com}

\altaffiltext{1}{Department of Physics and Astronomy, University of
California, Los Angeles, CA 90095-1562}
\altaffiltext{2}{Now at Ball Aerospace \& Technologies Corp., 
1600 Commerce St., M/S TT-2, Boulder, CO 80301}
\altaffiltext{3}{Harvard-Smithsonian Center for Astrophysics, 60
Garden Street, Cambridge, MA 02138}

\begin{abstract}
We have used the VLA to detect emission from the supergiant VY CMa at
radio wavelengths and have constructed 3000-4500 K isothermal outer
atmospheres constrained by the data.  These models produce a radio
photosphere at 1.5-2 R$_{\ast}$.  An extrapolation of the model can
account for the observed total mass-loss rate of the star. We also
present mid-infrared imaging of the supergiant that suggests warm dust
is extended in the same direction as the near-infrared reflection
nebula around VY CMa.  The origin of the asymmetries in the outflow
remains an unsolved problem.

\end{abstract}


\section{Introduction}

Massive stars (M$_{initial}$ $>$ 8-10 M$_\odot$) conclude their
stellar evolution as red supergiants before exploding as Type II
supernovae.  VY CMa (HD 58061; M5eIa) is one of the best studied
supergiants.  At a distance of 1.5 kpc \citep{lada78}, VY CMa
maintains a luminosity of L$_{\ast}$ = 2-5 $\times$ 10$^5$ L$_{\odot}$
\citep{jura90, lesidaner96, monnier99}, suggesting it is a massive
star (M $\ge$ 15 M$_{\odot}$) that will explode as a supernovae within
10$^5$ yr \citep{hirschi04}.  It is losing mass at a rate $\sim$ 2-4
$\times$ 10$^{-4}$ M$_{\odot}$ yr$^{-1}$ \citep{danchi94, monnier99}
which produces the optically thick, dusty envelope enshrouding the star,
thus hiding many clues to the nature of this enigmatic supergiant.

There are many mysteries regarding supernova physics including why a
core collapse event produces an explosion releasing $\sim$10$^{51}$
erg of kinetic energy and why some pulsars have space velocities as
high as 1000 km s$^{-1}$ \citep{strom95}.  Both of these effects may
be explained by anisotropies \citep{burrows00, khokhlov99,
macfadyen99}.  Indeed, many core collapse supernova remnants show
strong polarization (4\% for SN1997X; \citealt{wang01}), indicating a
high degree of asphericity.  

VY CMa exhibits asymmetries on scales from 10-10,000 R$_{\ast}$
(0$\farcs$1 - 100$\arcsec$) and, therefore, is a good candidate for
investigating the origin and nature of asymmetries in supergiants.
Early on, \citet{herbig72} showed that at optical wavelengths the
nebula surrounding VY CMa appeared irregularly shaped with apparent
structures within the bright central core of the nebulosity.  At large
scales ($\sim$ 3 pc), VY CMa's outflow is influencing the dynamics of
the nearby H II region S310 \citep{lada78}.  Recent visible and
near-infrared (near-IR) images of VY CMa from the Hubble Space
Telescope (HST) display a complex distribution of knots and filaments
embedded in a one-sided reflection nebula extending $\sim$ 4$\arcsec$
from the star \citep{smith01, kastner98} toward the southwest.  On
small scales, using near-IR aperture masking techniques,
\citet{monnier99, monnier04} have shown that the inner dust shell at
$\sim$ 15 R$_{\ast}$ is highly one-sided and consistent with the HST
results.  Even the molecular emission as traced by SiO is much
brighter and more extended to the southwest \citep{shinnaga03}.
Additionally, the spatial and redshift distribution of H$_2$O maser
emission is not consistent with a simple spherical outflow, but rather
appears elongated in the same direction as the near-IR emission
\citep{marvel96, richards98}.  However, the point-spread-function
(PSF) subtracted 8.4 and 9.8 $\micron$ images of VY CMa by
\citet{smith01} give the appearance of a bipolar structure within
2$\arcsec$ east and west of the central source in apparent disagreement 
with the other observations.  

Compared to the mass-loss envelope's structure, little is known about
the inner layers ($<$ 10 R$_{\ast}$) of VY CMa's atmosphere where the
dust condenses and mass is driven out due to radiation pressure on the
dust and subsequent drag on gas molecules.  Since the star is
concealed within an optically thick mass-loss envelope, only the
longest wavelengths can possibly probe the inner atmosphere of VY CMa.
\citet{knapp95} found that the 8.4 GHz emission measured by the VLA is
$\sim$ 6 times that expected from photospheric emission of the
supergiant and suggested that the excess may be due to an extended,
10,000 K ionized chromosphere as had been hypothesized for the nearby
(D $\sim$ 130 pc) supergiant $\alpha$ Ori \citep{newell82,
hartmann84}.  However, recent observations of $\alpha$ Ori revealed
its atmosphere is dominated by cool (few $\times$ 1000 K) gas
\citep{lim98}.  Further, \citet{lim98}, using the VLA in the
A-configuration, found that $\alpha$ Ori was resolved at 43 GHz with a
radius of $\sim$ 44 mas and partially resolved at 8.4, 15, and 22 GHz.
The radius at 43 GHz corresponds to a size of nearly twice the stellar
radius of 22.5 mas (3 AU; \citealt{dyck96}).  \citet{harper01}, using
the flux density and visibility measurements from various authors and
a modified version of the Mira radio photosphere model by
(\citealt{reid97}; hereafter RM97), created a semi-empirical model of
the temperature and density distributions out to $\sim$ 14 R$_{\ast}$
that fit the radio measurements of $\alpha$ Ori .

We have obtained VLA radio observations of VY CMa to test possible
emission scenarios and probe for asymmetries at radio frequencies.  We
find that a radio photosphere model similar to RM97 and
\citet{harper01} fit the observed data.  We use this model to
interpret the physical characteristics of the inner (1-10 R$_{\ast}$)
atmosphere of VY CMa.  We also present new mid-IR observations of VY
CMa at 11.7 and 17.9 $\micron$ and show that the thermal mid-IR
emission from the envelope is consistent with that seen in optical and
near-IR observations.

This paper is organized as follows: in $\S$ 2 we describe the
observations of VY CMa we have obtained with the VLA ($\S$ 2.1) and
the Keck Observatory ($\S$ 2.2).  We then discuss, in $\S$ 3, the
origin of the radio emission by first identifying emission models that
do not fit the data well and then in $\S$ 3.1 present an atmospheric
model containing a radio photosphere that reproduces the observed flux
densities.  A description of the analysis and discussion of our mid-IR
data is contained in $\S$ 4.  In $\S$ 5 we present our conclusions and
suggest possible future observations.

\section{Observations}
\subsection{VLA Radio Observations}
We observed VY CMa with the NRAO\footnote{The National Radio Astronomy 
Observatory is a facility of the National Science Foundation operated 
under cooperative agreement by Associated Universities, Inc.} Very
Large Array (VLA) on 2001 Oct. 2 at 43 GHz (Q-band) in the hybrid DnC
configuration and on 2002 Sept. 21 and 23 at 22 GHz (K-band) and 15
GHz (U-band) in the CnB configuration; see Table \ref{vla_obs} for
details.  We used the standard continuum observation technique,
employing four 50 MHz bands (right- and left-circular polarization at
two frequencies).  The central frequencies observed were 43315 and
43365 MHz, 22435 and 22485 MHz, and 14915 and 14965 MHz for Q-band,
K-band, and U-band, respectively.  Pointing was verified about every
hour.  For the 22 GHz and 15 GHz data, we employed fast-switching to
remove the tropospheric phase fluctuations.  The flux density scale
was set using observations of 3C 48 for the 43 GHz observations, and
3C 286 for the 22 GHz and 15 GHz observations.  The secondary
calibrator, 0650 - 166, was used to calibrate the interferometric
amplitudes and phases of VY CMa at all frequencies.  Table
\ref{vla_cal} lists the assumed calibration source flux densities.

Calibration and imaging of the visibilities followed the procedures
recommended for VLA data with the Astronomical Image Processing System
(AIPS).  Specifically, we imaged VY CMa at each wavelength with maps
of 256 by 256 pixels and pixel scales of 0$\farcs$14, 0$\farcs$1, and
0$\farcs$1 at 43, 22, and 15 GHz, respectively.  Due to the low 
declination of the source, we used hybrid VLA configurations with a 
a slightly longer northern arm which produced synthesized beams that 
are slightly elongated in the east-west direction: 1$\farcs$7 $\times$ 
1$\farcs$5 with PA =
-83$^{\circ}$ at 43 GHz; 1$\farcs$0 $\times$0$\farcs$78 with PA =
-70$^{\circ}$ at 22 GHz; and 1$\farcs$5 $\times$ 1$\farcs$0 with PA =
80$^{\circ}$ at 15 GHz.  At 43 and 15 GHz, VY CMa appears
marginally larger than the beam, with deconvolved geometric-mean sizes
of 1$\farcs$4 and 1$\farcs$7, respectively; at 22 GHz the source is
unresolved.  However, because the position angles for the elliptical
Gaussian fits to the sources do not match those of the beam, we
consider VY CMa to be unresolved at all three observed frequencies.

We list the peak flux density measurements in Table \ref{vla_vy}, with
formal errors that combine a 20\% systematic error and the RMS noise
error estimated from the background.  We note that the integrated flux
densities, from a Gaussian fit to the source, tend to be greater than
the peak flux densities by as much as 20\%.  This is likely the result
of poor phase calibration due to the $>$ 10$^{\circ}$ separation
between VY CMa and the secondary calibrator.  We consider the peak
flux density to be a better estimate of the flux since the source is
unresolved at each wavelength.

\subsection{Keck Mid-IR Observations}
We observed VY CMa with the Long Wavelength Spectrograph (LWS;
\citealt{jones93}), a facility instrument at the W.M. Keck
Observatory, on 2000 Feb. 5 at 11.7 $\micron$ ($\Delta \lambda$ = 1.0
$\micron$) and 17.9 $\micron$ ($\Delta \lambda$ = 2.0 $\micron$).  LWS
uses a 128 $\times$ 128 pixel Boeing Si:As detector and has a plate
scale of 0$\farcs$08 pixel$^{-1}$, resulting in a 10$\farcs$2 $\times$
10$\farcs$2 field of view.  We employed the ``chop-nod'' mode of
observing, with a chop throw of 10$\arcsec$ north.  The IRTF IR
standard $\alpha$ Aur (N = -1.94; Q = -1.93; \citealt{tokunaga84}) was
used to calibrate the flux of VY CMa and to measure the PSF.  The
full-width-at-half-maximum of $\alpha$ Aur was measured to be
0$\farcs$58 at 11.7 $\micron$ and 0$\farcs$48 at 17.9 $\micron$.

We measure fluxes of 8600$\pm$1700 Jy at 11.7 $\micron$ and
9300$\pm$1900 Jy at 17.9 $\micron$ for VY CMa.  The 20\% errors are
derived from the variability of the calibrator's flux throughout the
night.  Our measured flux density at 11.7 $\micron$ is consistent with
the color corrected \citep{iras88} IRAS 12 $\micron$ flux of 8729 Jy 
and with the mean DIRBE 12 $\micron$ flux of 9295.5 Jy.  

\section{Identifying the Origin of the Radio Emission}

As we have not resolved VY CMa at radio wavelengths, we cannot comment
on any asymmetry in the inner atmosphere, except to say that if one
exists, it must be on scales smaller than $\sim$ 1$\arcsec$ = 1500 AU.
The question we now ask is: what is the source of the radio emission
from VY CMa?

In Table \ref{vla_vy}, we have compiled the radio flux densities of VY
CMa in the frequency range 8.4 - 43 GHz from this work and the
literature.  From VLA monitoring of VY CMa at 8.4 GHz from September
1990 to May 1991, \citet{menten04} find that VY CMa may be variable by
about a factor of 2 on an irregular timescale, but they note that
within the errors, this variability is tentative.  \citet{knapp95}
measured a flux density of 0.26 mJy for VY CMa in January 1991,
consistent with the mean value of \citet{menten04}.  Our measurement
of VY CMa's flux density at 22 GHz of 0.66 $\pm$ 0.20 mJy is about 1/2
the 1.3 $\pm$ 0.3 mJy measured by \citet{menten04} in the early 1990's
at this frequency.  This suggests that VY CMa has some significant
variability at radio frequencies in addition to the irregular 
variability seen at
optical wavelengths \citep{wallerstein78} and the semi-regular 
variability seen in maser emission
\citep{harvey74, cox79, gomez-balboa86, martinez88}.  Further
observations will be necessary to confirm any variability.

Fitting a mean power law to the flux densities listed in Table
\ref{vla_vy} gives a spectral index of $\alpha$ $\sim$ 1.4 $\pm$ 0.2
(S$_{\nu}$ $\propto$ $\nu^{\alpha}$), but given the possible
variability, this value is uncertain since the phase at each
observation is unknown.  If we break up the data into the '1990's set'
and the '2000's set', where the former set is comprised of the
\citet{menten04} and \citet{knapp95} data and the latter of the data
presented herein, we find the spectral index of each set is 1.7 $\pm$
0.3, which is similar to that found for the radio spectra of Mira
variables ($\alpha$ = 1.86; RM97).  This spectral index is
also close that expected from the stellar photosphere ($\alpha$ = 2),
however, the flux density levels we measure are about twice that
expected from a star with photospheric temperature of T$_{\ast}$ =
2800 K and radius R$_{\ast}$ = 2.25 $\times$ 10$^{14}$ cm (15 AU;
\citealt{danchi94}) at the distance of VY CMa (see Figure \ref{spec},
solid line).

The sub-millimeter spectrum of VY CMa was modeled by
\citealt{knapp93}; hereafter KSR93) using small circumstellar dust
grains (a$_g$ = 2000 \AA; $\rho_g$ = 3.5 g cm$^{-3}$) with an
emissivity index $\beta$ = 0.9.  They derive a dust mass-loss rate of
2 $\times$ 10$^{-6}$ M$_\odot$ yr$^{-1}$ within a region of radius
10$^{18}$ cm.  Assuming all the emission above the expected
photospheric emission at radio frequencies is due to dust with the
KSR93 parameters, then we calculate the total dust mass using
\begin{equation}
M_{dust} =\frac{2 (S_{\nu}-S_{phot}) a_g \rho_g D^2 c^2}{3 Q_{\nu} k_b
T_{env} \nu^2}
\end{equation}
where $S_{\nu}$ is the total measured flux density from the source,
$S_{phot}$ is the expected photospheric flux density, $D$
is the distance to VY CMa, $Q_{\nu}$ is the grain emissivity, and
$T_{env}$ is the temperature in the envelope.  We use the KSR93
function for the emissivity: $Q_{\nu} = Q_0(\nu/\nu_0)^\beta$ where
$Q_0$ = 5.65 $\times$ 10$^{-4}$ at $\nu_0$ = 274.6 GHz.  The color
corrected IRAS fluxes \citep{iras88} at 60 and 100 $\micron$
are 1119 Jy and 304.5 Jy, respectively, and the ratio of the IRAS
fluxes gives an envelope temperature of 235 K.  Thus, our 43.3 GHz
flux density of 2.8 mJy, after subtraction of 1.2 mJy from the
expected photospheric emission, implies a dust mass of 1.3 $\times$
10$^{-3}$ M$_{\odot}$ within our unresolved source of radius $\sim$
0$\farcs$5.  However, using the dust mass-loss rate and outflow
velocity from the KSR93 model, the total mass expected to reside
within a radius of 0$\farcs$5 is only 1.8 $\times$ 10$^{-4}$
M$_{\odot}$, a factor of $\sim$ 7 less than that calculated from the
43.3 GHz data.  That is, there would need to be more dust closer to
the star than is predicted by the KSR93 model.  It may be possible to
fit the observed radio flux densities with a second population of dust
grains closer to the star.

\subsection{Radio Photosphere Model}
We now turn to an alternative source of radio emission: a radio
photosphere.  In their attempt to model radio spectra of Mira stars,
RM97 find that at long wavelengths their observations of these long
period variables are well fit by radio photospheres with radii $\sim$
2 R$_{\ast}$.  The opacity in the extended radio photosphere model is
predominately provided by free-free absorption from the interaction
between neutral atomic and molecular hydrogen and free electrons
supplied mostly by Na, Al, K, and Ca (elements with low ionization
potentials).  We have applied this opacity model to VY CMa.

Given a temperature and density profile for the stellar atmosphere as
a function of radius outward from the surface, at each layer in the
star's atmosphere we assume thermodynamic equilibrium and solve the
Saha equation simultaneously for Na, Al, K, and Ca to determine the
total electron density and pressure (P$_e$).  Following RM97, we use
solar abundances, ionization potentials, and partition functions from
\citet{gray92}.  We use dissociation coefficients given by
\citet{tsuji64} to calculate the densities of atomic and molecular
hydrogen ($\rho$($H^-$) and $\rho$(H$_{2}^-$)).  To evaluate the
absorption coefficients for H$^-$ and H$_{2}^-$ free-free emission
($\kappa_{\nu}$(H$^-$) and $\kappa_{\nu}$(H$_{2}^-$)) for temperatures
$<$ 3000 K, we use the third order polynomials in temperature given in
RM97.  For temperatures higher than 3000 K, we assume the atmosphere
is composed entirely of atomic hydrogen, and solve for the absorption
coefficient, $\kappa_{\nu}$(H$^-$), by numerically evaluating the
equations given in \citet{dalgarno66}.  The absorption coefficients
are proportional to the inverse square of the frequency and have units
of cm$^4$ dyn$^{-1}$.  We then calculate the optical depth increment
\begin{equation}
d\tau = [\kappa_{\nu}(H^-) \rho(H^-) + \kappa_{\nu}(H_2^-)
\rho(H_2^-)] P_e dl
\end{equation}
where $dl$ is the step size through the atmosphere.  The brightness
temperature, $T_b$, at an optical depth $\tau_0$ in the supergiant's
atmosphere is computed from
\begin{equation}
T_b(\tau_0) = T(\tau_0)e^{-\tau_0} + \int_{0}^{\tau_0}
T(\tau)e^{\tau-\tau_0}d\tau 
\end{equation}
which is the formal solution to the radiative transfer equation in the
Rayleigh-Jeans limit.  We numerically integrate along the
line-of-sight toward the center of the star and then repeat the
integrations moving perpendicular to the line-of-sight with step size
$\delta p$ to evaluate the center-to-limb brightness temperature
profile, $T_b(p)$.  The total flux density, S$_{\nu}$, is
calculated assuming azimuthal symmetry using
\begin{equation}
S_{\nu} = \sum_{p=0} \frac{T_b(\tau_0,p) 2 \pi p \delta p}{D^2}
\end{equation} 
We use a model's total flux density values at all four observed radio
frequencies compared to those observed to evaluate the validity of a
model and its respective inputs result.

As a simple approximation, we assume an isothermal atmosphere and
compute the density structure for a range of input temperatures.  The
density profile for an isothermal atmosphere is that required by
hydrostatic equilibrium out to the critical radius ($r_{crit} = G
\epsilon M_{\ast}/2 a^2$, where $a = (k_b T/\mu)^{1/2}$ is the
isothermal sound speed, $G$ is the gravitational constant, $\epsilon
M_{\ast}$ is the effective stellar mass, and $\mu$ is the mean
molecular weight of the gas) even if there is a radiation pressure
driven wind \citep{lc99}.  We have used the term $\epsilon = (1 -
L_{\ast} \chi/4 \pi c G M_{\ast})$, where $\chi$ is the opacity, to
account for the alteration of the effective stellar mass due to
radiation pressure.  We take the temperature- and density-dependent
Planck mean opacity values from \citet{helling00} as upper limits to
the opacity and consider this to be more accurate for the optically
thin stellar atmosphere than the Rosseland mean opacities
\citep{hofner98}.  At temperatures higher than $\sim$ 3000 K the
Rosseland mean opacities are essentially equivalent to the Planck mean
opacities \citep{helling00}. The density as a function of radius, $r$,
for an atmosphere in hydrostatic equilibrium can be computed from
\begin{equation}
\rho(r) = \rho(R_{\ast}) ~exp\displaystyle\bigg\{
-\frac{(r-R_{\ast})}{H_{\ast}}\frac{R_{\ast}}{r}\displaystyle\bigg\}
\end{equation}
where $\rho(R_{\ast})$ is the density at R$_{\ast}$, and $H_{\ast} =
k_b T_{\ast}/\mu g_{\ast}$ is the scale height with $g_{\ast} = G
\epsilon M_{\ast}/R_{\ast}^2$.

We use a stellar radius of R$_{\ast}$ = 2.25 $\times$ 10$^{14}$ cm, an
initial density of $\rho_{\ast}$ = 5 $\times$ 10$^{14}$ cm$^{-3}$, and
a stellar mass of M$_{\ast}$ = 15 M$_{\odot}$.  The initial density
has been estimated by assuming a 2800 K optical photosphere in
hydrostatic equilibrium with a Rosseland mean opacity of $\chi$ =
10$^{-4}$ cm$^2$ g$^{-1}$ at an optical depth of $\tau$ = 2/3.  We
report the results of three isothermal model atmospheres: T = 1600 K,
3000 K, and 4500 K.  Figure \ref{iso_td} shows the density profile for
each model while Figure \ref{iso_tb} shows the output brightness
temperature as a function of impact parameter for each model at 22
GHz.

For the T = 1600 K model atmosphere, the density drops so quickly that
the optical depth toward impact parameters $>$ R$_{\ast}$ is very
small and, therefore, the radius of the radio star is not much larger
than R$_{\ast}$.  Since S$_{\nu}$ is a sum over the radio radius, this
results in flux density values too low to be consistent with those
measured for VY CMa.

We find that the two warmer atmospheres produce radio photospheres
whose model flux densities bracket those observed for VY CMa.  The
model output flux density values from the radio photosphere for both
these models are listed in Table \ref{vla_vy} and are represented as
the dashed and dot-dashed lines in Figure \ref{spec}.  The spectral
index for both models is 1.8.  Because of the optically thick envelope
enshrouding VY CMa, the supergiant lacks any UV lines which would hint
at the existence and temperature of a stellar chromosphere.  Recent
modeling of high resolution UV spectra of the supergiant $\alpha$
Ori's atmosphere produced a temperature fit to the H$\alpha$ lines not
in excess of 5500 K \citep{lobel03}.  By analogy, such a distribution
of warm plasma may exist around VY CMa.

Using these models, we find that the apparent radius at radio
frequencies, estimated as the half-width at half-brightness of the
$T_b(p)$ profile, is 1.5-2 R$_{\ast}$ and varies inversely with
frequency; see Figure \ref{iso_rad}. The dependence of apparent radius
on frequency is consistent with the pattern observed by \citet{lim98}
for the resolved and partially resolved radii of $\alpha$ Ori.  The
apparent radius of VY CMa's model radio photosphere is comparable to
the 2 R$_{\ast}$ radio photosphere proposed for Mira stars (RM97) and
is consistent with the radio emitting region fit by semi-empirical
models in \citet{harper01} for $\alpha$ Ori (see their Figure 5).

The mass-loss rate for an isothermal atmosphere can be calculated from
\begin{equation}
\dot{M}_{tot} = 4 \pi \rho_{\ast} a R_{\ast}^2
\displaystyle\bigg\{\frac{v_{esc}(R_{\ast})}{2
a}\displaystyle\bigg\}^4
exp\displaystyle\bigg\{-\frac{v^2_{esc}(R_{\ast})}{2 a^2} +
\frac{3}{2}\displaystyle\bigg\}
\end{equation}
(corrected from \citealt{lc99}), where $v_{esc}(R_{\ast}) =
(2~G~\epsilon M_{\ast}/R_{\ast})^{1/2}$ is the escape velocity at the
stellar photosphere.  For the T = 3000 K model, then $a$ = 5.0 km
s$^{-1}$ and $v_{esc}(R_{\ast})$ = 42.1 km s$^{-1}$, which gives
$\dot{M}_{tot}$ = 1.9 $\times$ 10$^{-12}$ M$_{\odot}$ yr$^{-1}$, about
8 orders of magnitude lower than what has been measured for VY CMa.
For the T = 4500 K model, however, $a$ = 6.1 km s$^{-1}$ and
$v_{esc}(R_{\ast})$ = 33.9 km s$^{-1}$, which gives $\dot{M}_{tot}$ =
2.8 $\times$ 10$^{-4}$ M$_{\odot}$ yr$^{-1}$, comparable to the
value measured by molecular line emission of 2-4 $\times$ 10$^{-4}$ 
M$_{\odot}$ yr$^{-1}$ (see \citet{danchi94} for a summary).

We have also applied the radio photosphere model to an adiabatic
atmosphere.  However, with an adiabatic atmosphere in hydrostatic
equilibrium, we find that the optical depth due to the H$^-$ and
H$_{2}^{-}$ opacity never reaches $>$ 1.  At high temperatures, where
the Planck mean opacity is large ($\chi$ = 0.32 for
T$_{\ast}$ = 6000 K), for the atmosphere to be in hydrostatic
equilibrium the density as a function of radius must drop faster
compared to lower temperatures.  Once the density falls below a
threshold density (n$_H$ $\sim$ 10$^{12}$ cm$^{-3}$), there simply are
not enough free electrons for the H$^-$ and H$_{2}^-$ opacity to be
effective.

\section{Mid-IR Emission}
Our new resolved mid-IR images of VY CMa with the LWS at the
W. M. Keck Observatory are shown in Figures \ref{mir}a \& c.  We find
that VY CMa is asymmetrically extended to the southwest at both 11.7
and 17.9 $\micron$ with the maximum extension at position angle
277$^{\circ}$ as measured East from North.  Figure \ref{mir}b \& d
show the radial profiles at maximum extension (277$^{\circ}$), minimum
extension (47$^{\circ}$), and at the normal to the maximum extension
(137$^{\circ}$ \& 317$^{\circ}$) compared to the azimuthally averaged
radial profile of the standard star $\alpha$ Aur (solid line with
error bars).  We find that at 11.7 (17.9) $\micron$, at a distance of
0$\farcs$8, the flux is 3.4 (2.0) times greater toward position angle
227$^{\circ}$ compared to along position angle 47$^{\circ}$.  At a
distance of 1$\farcs$0, the ratio of fluxes is 3.4 (3.5) at 11.7
(17.9) $\micron$.  The direction of the extended emission is
consistent with that seen in the optical and near-IR.  We find no
evidence for the bi-polar east-west structure surrounding the central
star at a distance of $\sim$ 2$\arcsec$ as seen in the PSF-subtracted
images of \citet{smith01}.  The average polarization of the SiO masers
in the $v$ = 1, $J$ = 5-4 transition, 251$^{\circ}$, is essentially
aligned with the maximum extent of the mid-IR emission
\citep{shinnaga04}.  This implies that the asymmetrical envelope is
not just caused by a scattering effect, but that the mass itself is
anisotropically distributed, perhaps by a bipolar outflow.

The color temperature of the dust grains is estimated to be $\sim$ 245
K from the ratio F$_{nu}$(11.7 $\micron$)/F$_{nu}$(17.9 $\micron$) =
0.92.  We estimate the dust mass $M_{dust}$ using
\begin{equation}
M_{dust} = \frac{F_{\nu} D^2}{\chi_{\nu} B_{\nu}(T_g)}
\end{equation}
where $F_{\nu}$ is the observed flux at 17.9 $\micron$ and
$\chi_{\nu}$ is the grain opacity.  With T$_d$ = 245K, and
$\chi_{\nu}$(17.9 $\micron$) $\sim$ 600 cm g$^{-1}$ \citep{sopka85},
we find M$_{dust}$ = 5.8 $\times$ 10$^{-4}$ M$_{\odot}$.  This is a
lower limit to the dust mass since we have assumed the circumstellar
envelope is optically thin in this calculation.  The dust mass-loss
rate (\.{M}$_{dust}$) can be estimated using the approximate size of
the dust emitting region (r$_{dust}$ $\sim$ 2$\arcsec$) and the
outflow velocity ($v_{out}$ = 39 km s$^{-1}$; \citealt{sopka85}) and
is found to be 1.6 $\times$ 10$^{-6}$ M$_{\odot}$ yr$^{-1}$, which is
consistent with the value obtained from 400 $\micron$ data by
\citet{sopka85} of 7.8 $\times$ 10$^{-7}$ M$_{\odot}$ yr$^{-1}$ and
the upper limit of 3.5 $\times$ 10$^{-6}$ M$_{\odot}$ yr$^{-1}$
published by KSR93 and calculated from 1.1 mm data.

\section{Conclusions} 
We report the detection of emission from the supergiant VY CMa at
radio frequencies with the VLA.  We have constructed isothermal outer
atmospheres of 3000-4500 K containing radio photospheres where
interactions between free electrons and neutral and molecular hydrogen
provide the dominant opacity.  These isothermal atmospheric models
produce a radio photosphere at 1.5-2 R$_{\ast}$ and flux density
values and a spectral index comparable to those measured for VY CMa.
An extrapolation of the model can account for the observed total
mass-loss rate observed for VY CMa.  We also present resolved mid-IR
imaging of VY CMa.  These data suggest that warm dust is extended in
the same direction as the near-IR reflection nebula around VY CMa and
that the mass-loss rate as derived from the mid-IR is consistent with
previous measurements.

In their VLA observations of $\alpha$ Ori, \citet{lim98} resolved the
supergiant's radio surface at 43 GHz and found that the star was
asymmetrically shaped at this frequency.  The authors proposed that
such an asymmetry may arise from giant convection cells elevating cool
material into the atmosphere.  Our VLA radio observations at 15, 22,
and 43 GHz produce unresolved images of VY CMa and thus we cannot
comment on the asymmetry in the supergiant's inner atmosphere.  To
resolve a radio photosphere at 1.5-2 R$_{\ast}$ would require a
resolution of $\sim$ 0$\farcs$04 which is just beyond the capability
of the VLA at 43 GHz.  However, it will be possible to resolve VY CMa
at higher frequencies with the Atacama Large Millimeter Array (ALMA).
We expect a flux density at 345 GHz of $\sim$ 120-240 mJy with an
apparent radius of $\sim$ 0$\farcs$025, which would be easily resolved
with ALMA's largest D = 10 km configuration.  A resolved detection
would confirm the existence of a radio photosphere in VY CMa and may
provide clues to the symmetry of such an extended photosphere.
Continued monitoring of VY CMa at radio frequencies will help identify
any variability.  Further, monitoring of VY CMa in the IR as proposed
by \citet{monnier04} and \citet{smith01} to follow evolution of
structures and the general shape of the mass-loss envelope will
provide insight into the orientation and dynamics of the central star
and may help constrain the dust mass-loss parameters.

\acknowledgments This work has been supported by funding from NASA.  
The authors wish to thank Jean Turner and James Larkin for their 
insightful comments and Christine Chen for helping to obtain the 
Keck data.  
Data presented herein were obtained at the W.M. Keck Observatory,
which is operated as a scientific partnership among the California
Institute of Technology, the University of California and NASA. The
Observatory was made possible by the generous financial support of the
W.M. Keck Foundation.  The authors wish to recognize and acknowledge
the very significant cultural role that the summit of Mauna Kea has
always had within the indigenous Hawaiian community.  We are most
fortunate to have the opportunity to conduct observations from this
mountain.

\begin{deluxetable}{cccccc}
\tablecaption{VLA Observations of VY CMa\label{vla_obs}}
\tablewidth{0pt}
\tablehead{
\colhead{Date} & \colhead{Freq.} & \colhead{Config.} & 
\colhead{On-Source Time} & \colhead{Beam Size} & \colhead{PA}\\
\colhead{}     & \colhead{(GHz)} & \colhead{}        &
\colhead{(hr)}           & \colhead{}          & \colhead{(deg)}
}
\startdata
2001 Oct. 2 & 43 & DnC & 3.2 & 1$\farcs$7 $\times$ 1$\farcs$5 & -83 \\
2002 Sept. 21 \& 23 & 22 & CnB & 3.3 & 1$\farcs$0 $\times$ 0$\farcs$8 & -70 \\
2002 Sept. 21 \& 23 & 15 & CnB & 3.3 & 1$\farcs$5 $\times$ 1$\farcs$2 & 80 \\
\enddata
\end{deluxetable}

\begin{deluxetable}{cccccc}
\tablecaption{Calibrators\label{vla_cal}}
\tablewidth{0pt}
\tablehead{
\colhead{Source} & \colhead{R.A.} & \colhead{Dec.} & 
\colhead{F$_{15 GHz}$} & \colhead{F$_{22 GHz}$} & \colhead{F$_{43 GHz}$} \\
\colhead{}       & \colhead{(J2000)}            & \colhead{(J2000)} &
\colhead{(Jy)}       & \colhead{(Jy)}       & \colhead{(Jy)} 
}
\startdata
3C48  & 01 37 41.2999 & +33 09 35.133 & & & 0.53 \\
3C286 & 13 31 08.2880 & +30 30 32.959 & 3.45 & 2.52 & \\
0650-166 & 06 50 24.5819 & -16 37 39.725 & 2.55 & 2.38 & 1.18 \\
\enddata
\end{deluxetable}

\begin{deluxetable}{cccc}
\tablecaption{Observed and Model Flux Densities for VY CMa \label{vla_vy}}
\tablewidth{0pt}
\tablehead{
\colhead{Frequency} & \colhead{Measured F$_{\nu}$ } & 
\colhead{F$_{\nu}$(T = 3000 K Model) } &
\colhead{F$_{\nu}$(T = 4500 K Model) } \\
\colhead{(GHz)} & \colhead{(mJy)} & \colhead{(mJy)} & \colhead{(mJy)} 
}
\startdata
8.4 & 0.26 $\pm$ 0.03 \tablenotemark{a}; 0.24 $\pm$ 0.02 \tablenotemark{b} & 0.10 & 0.24 \\
15  & 0.41 $\pm$ 0.092 \tablenotemark{c} & 0.29 & 0.70 \\
22  & 1.3  $\pm$ 0.3 \tablenotemark{b}; 0.69 $\pm$ 0.15 \tablenotemark{c} & 0.61 & 1.4 \\
43  & 2.5  $\pm$ 0.51 \tablenotemark{c} & 2.2 & 4.9 \\
\enddata
\tablenotetext{a}{\citet{knapp95}; data from January 1991}
\tablenotetext{b}{\citet{menten04}; data from September 1990 to May 1991}
\tablenotetext{c}{This work}
\end{deluxetable}


\figcaption[vycma_bbdy_chrom.ps]{Measured flux densities of VY CMa
from 8.4 to 43 GHz from Table \ref{vla_vy}.  The solid line represents
a blackbody curve for a 2800K star with radius 2.25 $\times$ 10$^{13}$
cm.  The dashed (dotted) line connects the output flux densities from
the 3000 K (4500 K) isothermal atmospheric model with a radio
photosphere (see $\S$ 3.1 for details). \label{spec}}

\figcaption[iso_td.ps]{ The density profiles for isothermal stellar
atmospheres with R$_{\ast}$ = 2.25 $\times$ 10$^{14}$ cm,
$\rho_{\ast}$ = 5 $\times$ 10$^{14}$ cm$^{-3}$, and M$_{\ast}$ = 15
M$_{\odot}$.  Dot-dashed line: T = 1600 K; dashed line: T = 3000 K;
dotted line: T = 4500 K.\label{iso_td} }

\figcaption[iso_tb.ps]{Calculated brightness temperature from H$^-$
and H$_2^-$ free-free interactions at a frequency of 22 GHz for the
isothermal stellar atmosphere models in Figure \ref{iso_td}.
\label{iso_tb}}

\figcaption[freq_rad.ps]{Apparent stellar radius at radio wavelengths
as a function of frequency for both 3000 K (dashed line) and 4500 K
(dotted line) isothermal radio photosphere models described in the
paper and depicted in Figures \ref{iso_td} \& \ref{iso_tb}.
\label{iso_rad}}

\figcaption[vyfig.ps]{a) The 11.7 $\micron$ image of VY CMa.  North is
up and East to the left.  b) Cuts through the 11.7 $\micron$ image at
angles representing the semi-major axis of the extended source
(227$^{\circ}$ and 47$^{\circ}$) and the semi-minor axis
(137$^{\circ}$ and 317$^{\circ}$) compared to the azimuthally averaged
PSF of the standard star $\alpha$ Aur (solid line with error bars).
The angles are measured East from North. c) the 17.9 $\micron$ image
of VY CMa.  d) same as b) for the 17.9 $\micron$ data. \label{mir}}

\begin{figure}
\plotone{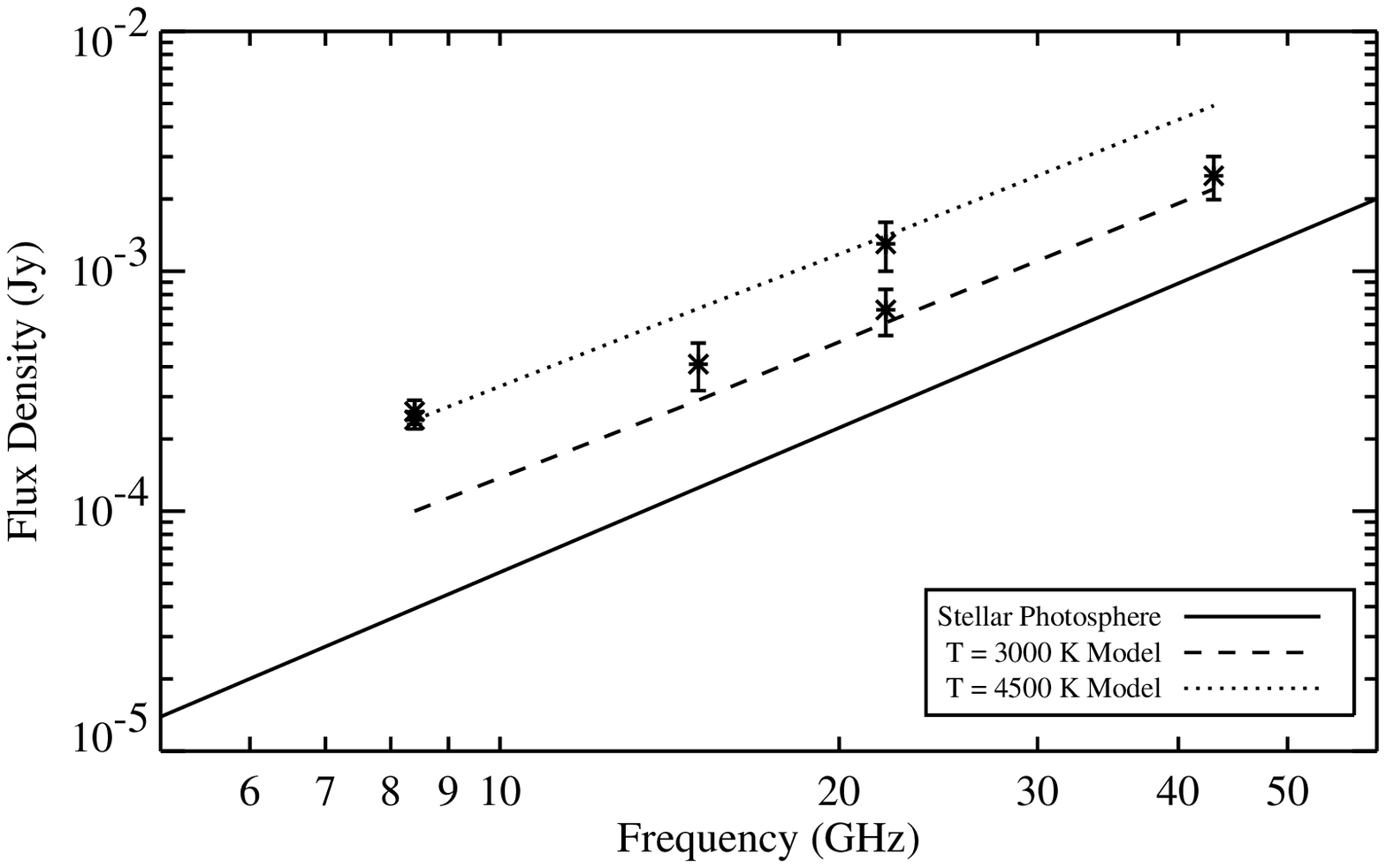}
\end{figure}

\begin{figure}
\plotone{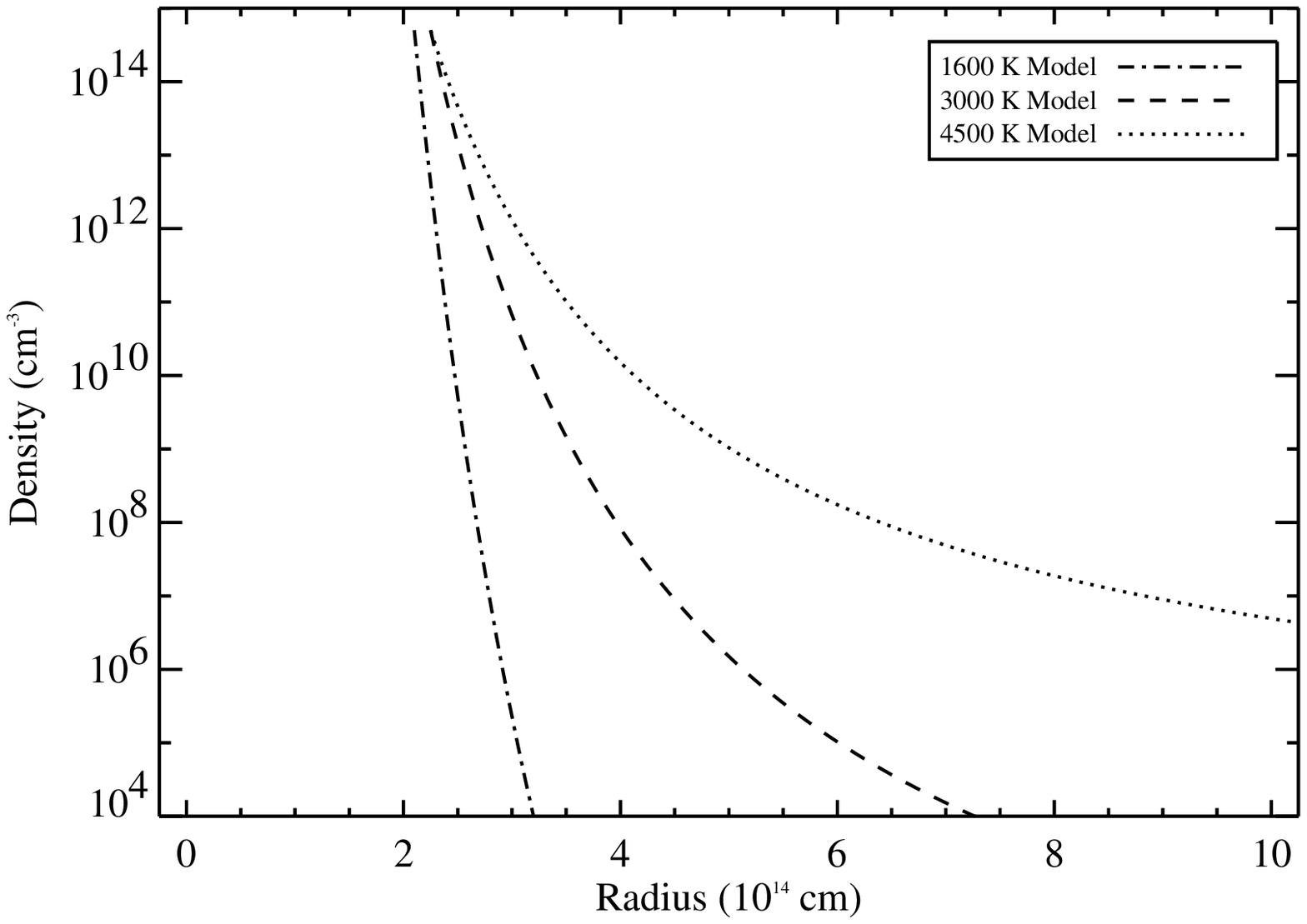}
\end{figure}

\begin{figure}
\plotone{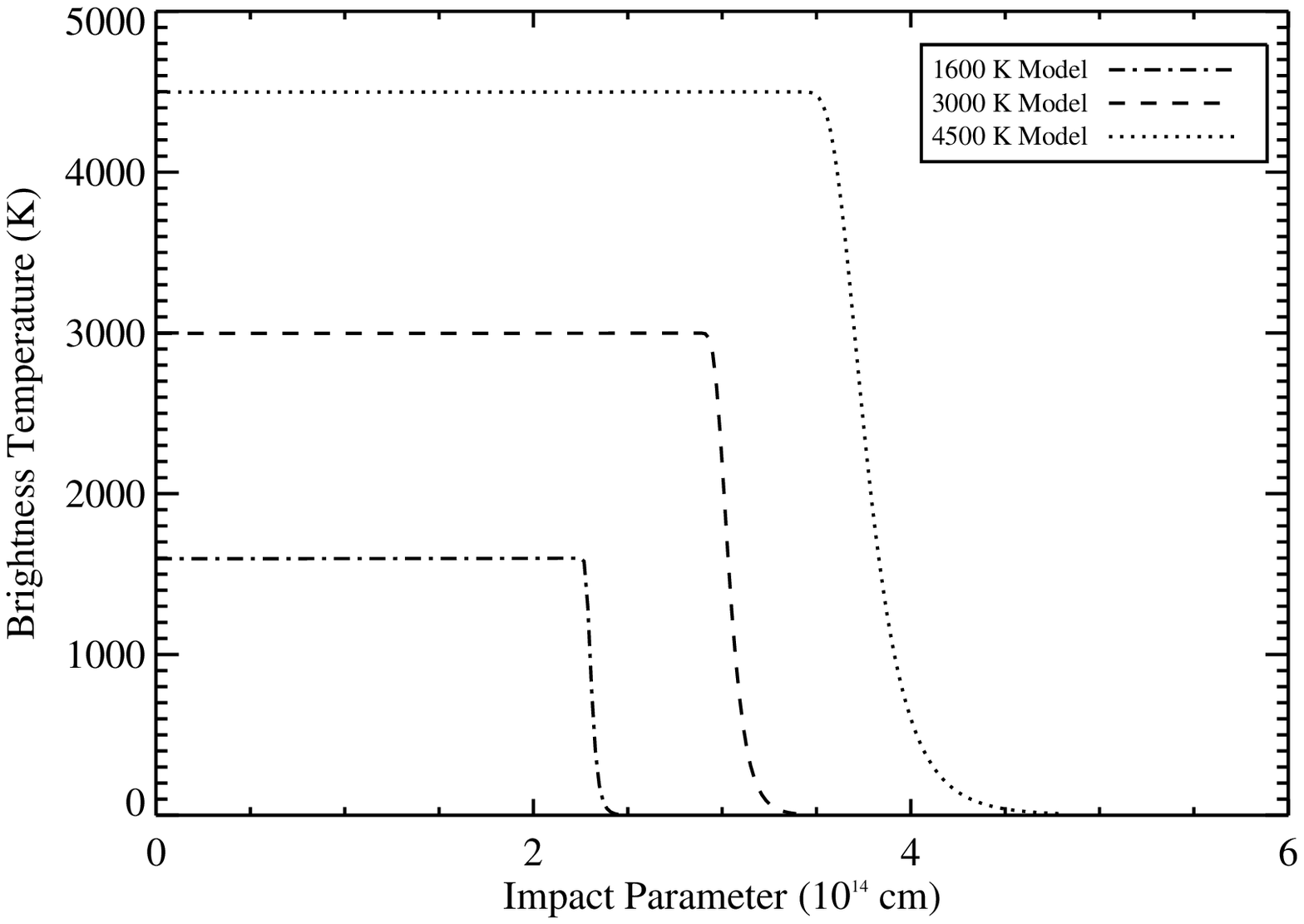}
\end{figure}

\begin{figure}
\plotone{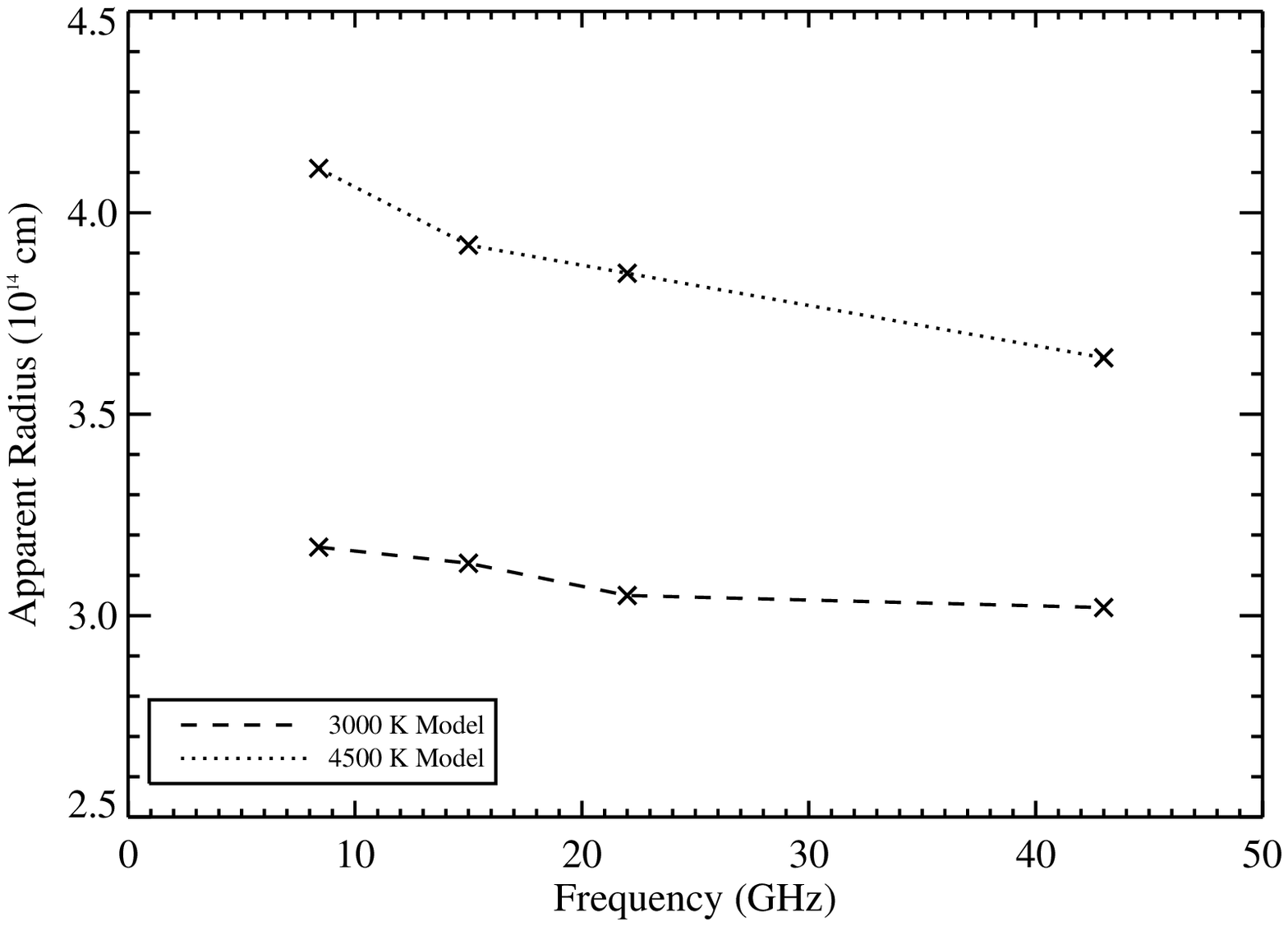}
\end{figure}

\begin{figure}
\plotone{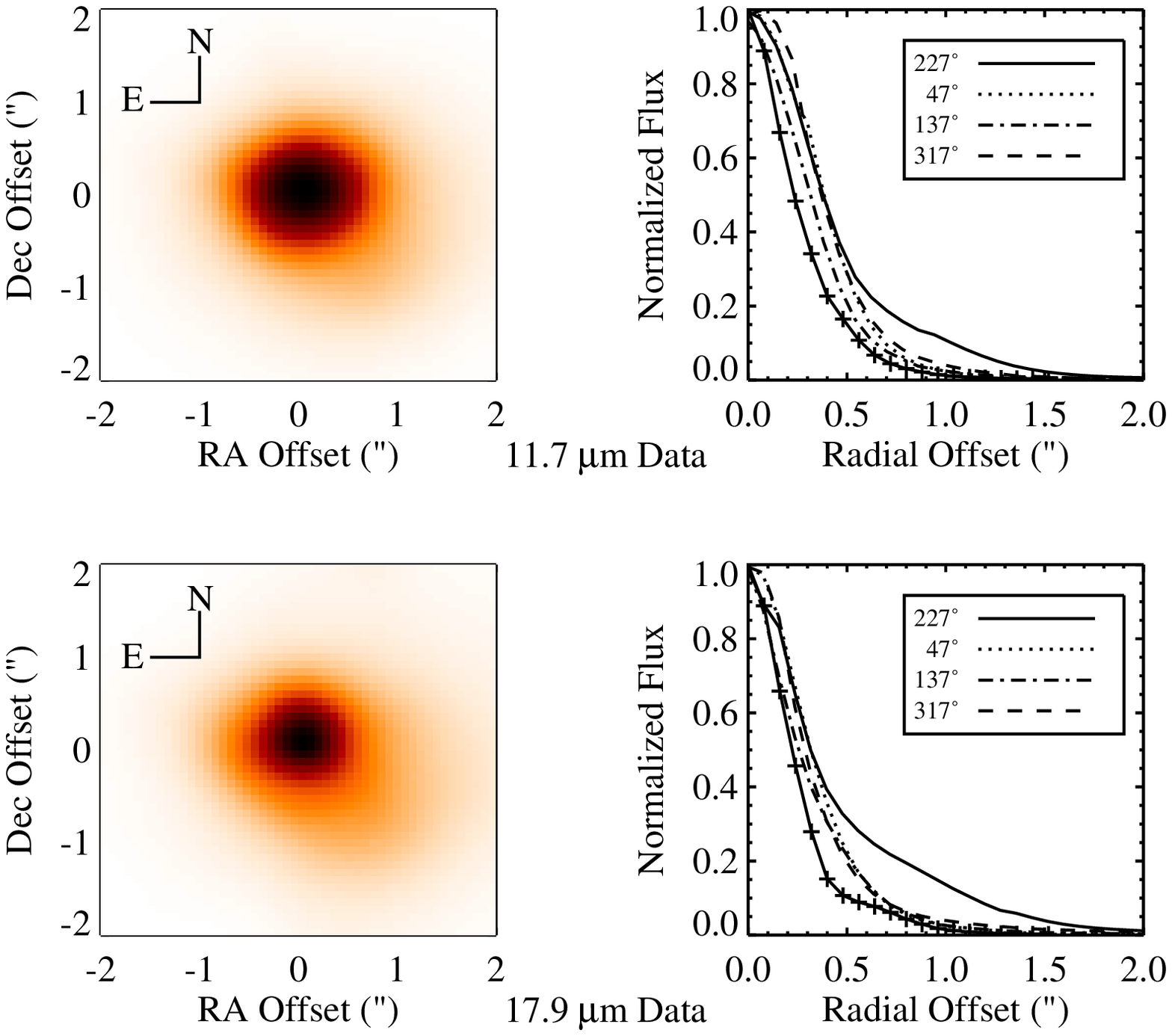}
\end{figure}


\begin{thebibliography}{}
  
\bibitem[Beichman et al.(1988)]{iras88} Beichman, C.~A., 
Neugebauer, G., Habing, H.~J., Clegg, P.~E., \& Chester, T.~J.\ 1988, NASA 
RP-1190, Vol.~1 (1988), 0 
\bibitem[Burrows(2000)]{burrows00} Burrows, A.\ 2000, \nat, 403, 727 
\bibitem[Cox \& Parker(1979)]{cox79} Cox, G.~G.~\& Parker, 
E.~A.\ 1979, \mnras, 186, 197 
\bibitem[Dalgarno \& Lane(1966)]{dalgarno66} Dalgarno, A.~\& Lane, 
N.~F.\ 1966, \apj, 145, 623 
\bibitem[Danchi et al.(1994)]{danchi94} Danchi, W.~C., 
Bester, M., Degiacomi, C.~G., Greenhill, L.~J., \& Townes, C.~H.\ 1994, 
\aj, 107, 1469 
\bibitem[Dyck, Benson, van Belle, \& Ridgway(1996)]{dyck96} 
Dyck, H.~M., Benson, J.~A., van Belle, G.~T., \& Ridgway, S.~T.\ 1996, \aj, 
111, 1705 
\bibitem[Gomez Balboa \& Lepine(1986)]{gomez-balboa86} Gomez Balboa, 
A.~M.~\& Lepine, J.~R.~D.\ 1986, \aap, 159, 166 
\bibitem[Gray(1992)]{gray92} Gray, D.~F.\ 1992, Cambridge 
Astrophysics Series, Cambridge: Cambridge University Press, 1992, 2nd ed., 
ISBN 0521403200.
\bibitem[Harper, Brown, \& Lim(2001)]{harper01} Harper, G.~M., 
Brown, A., \& Lim, J.\ 2001, \apj, 551, 1073 
\bibitem[Hartmann \& Avrett(1984)]{hartmann84} Hartmann, L.~\& 
Avrett, E.~H.\ 1984, \apj, 284, 238 
\bibitem[Harvey, Bechis, Wilson, \& Ball(1974)]{harvey74} 
Harvey, P.~M., Bechis, K.~P., Wilson, W.~J., \& Ball, J.~A.\ 1974, \apjs, 
27, 331 
\bibitem[Helling, Winters, \& Sedlmayr(2000)]{helling00} Helling, 
C., Winters, J.~M., \& Sedlmayr, E.\ 2000, \aap, 358, 651 
\bibitem[Herbig(1972)]{herbig72} Herbig, G.~H.\ 1972, \apj, 172, 
375
\bibitem[Hirschi et al.(2004)]{hirschi04} Hirschi, R., Meynet, 
G., \& Maeder, A.\ 2004, \aap, 425, 649 
\bibitem[Hoefner, Jorgensen, Loidl, \& Aringer(1998)]{hofner98} 
Hoefner, S., Jorgensen, U.~G., Loidl, R., \& Aringer, B.\ 1998, \aap, 340, 
497 
\bibitem[Jones \& Puetter(1993)]{jones93} Jones, B.~\& Puetter, 
R.~C.\ 1993, \procspie, 1946, 610 
\bibitem[Jura \& Kleinmann(1990)]{jura90} Jura, M.~\& 
Kleinmann, S.~G.\ 1990, \apjs, 73, 769 
\bibitem[Kastner \& Weintraub(1998)]{kastner98} Kastner, J.~H.~\& 
Weintraub, D.~A.\ 1998, \aj, 115, 1592 
\bibitem[Khokhlov et al.(1999)]{khokhlov99} Khokhlov, A.~M., H{\" 
o}flich, P.~A., Oran, E.~S., Wheeler, J.~C., Wang, L., \& Chtchelkanova, 
A.~Y.\ 1999, \apjl, 524, L107 
\bibitem[Knapp, Bowers, Young, \& Phillips(1995)]{knapp95} 
Knapp, G.~R., Bowers, P.~F., Young, K., \& Phillips, T.~G.\ 1995, \apj, 
455, 293 
\bibitem[Knapp, Sandell, \& Robson(1993)]{knapp93} Knapp, 
G.~R., Sandell, G., \& Robson, E.~I.\ 1993, \apjs, 88, 173 (KSR93)
\bibitem[Lada \& Reid(1978)]{lada78} Lada, C.~J.~\& Reid, 
M.~J.\ 1978, \apj, 219, 95 
\bibitem[Lamers \& Cassinelli(1999)]{lc99} Lamers, 
H.~J.~G.~L.~M.~\& Cassinelli, J.~P.\ 1999, Introduction to stellar winds / 
Henny J.G.L.M.~Lamers and Joseph P.~Cassinelli.~Cambridge ; New York : 
Cambridge University Press, 1999.~ ISBN 0521593980
\bibitem[Le Sidaner \& Le Bertre(1996)]{lesidaner96} Le Sidaner, 
P.~\& Le Bertre, T.\ 1996, \aap, 314, 896 
\bibitem[Lim et al.(1998)]{lim98} Lim, J., Carilli, C.~L., 
White, S.~M., Beasley, A.~J., \& Marson, R.~G.\ 1998, \nat, 392, 575 
\bibitem[Lobel(2003)]{lobel03} Lobel, A.\ 2003, The Future of 
Cool-Star Astrophysics: 12th Cambridge Workshop on Cool Stars , Stellar 
Systems, and the Sun (2001 July 30 - August 3), eds.~A.~Brown, G.M.~Harper, 
and T.R.~Ayres, (University of Colorado), 2003, p.~329-343., 12, 329 
\bibitem[Martinez, Bujarrabal, \& Alcolea(1988)]{martinez88} 
Martinez, A., Bujarrabal, V., \& Alcolea, J.\ 1988, \aaps, 74, 273 
\bibitem[Marvel(1996)]{marvel96} Marvel, K.~B.\ 1996, 
Ph.D.~Thesis 
\bibitem[MacFadyen \& Woosley(1999)]{macfadyen99} MacFadyen, 
A.~I.~\& Woosley, S.~E.\ 1999, \apj, 524, 262 
\bibitem[Menten \& Reid(2004)]{menten04} Menten, K.~M. \& Reid, M.~J., 
in prep. 
\bibitem[Monnier et al.(2004)]{monnier04} Monnier, J.~D., et al.\ 
2004, \apj, 605, 436
\bibitem[Monnier et al.(1999)]{monnier99} Monnier, J.~D., 
Tuthill, P.~G., Lopez, B., Cruzalebes, P., Danchi, W.~C., \& Haniff, C.~A.\ 
1999, \apj, 512, 351 
\bibitem[Newell \& Hjellming(1982)]{newell82} 
Newell, R.~T.~\& Hjellming, R.~M.\ 1982, \apjl, 263, L85 
\bibitem[Reid \& Menten(1997)]{reid97} Reid, M.~J.~\& Menten, 
K.~M.\ 1997, \apj, 476, 327 (RM97)
\bibitem[Richards, Yates, \& Cohen(1998)]{richards98} Richards, 
A.~M.~S., Yates, J.~A., \& Cohen, R.~J.\ 1998, \mnras, 299, 319 
\bibitem[Shinnaga, Moran, Young \& Ho(2003)]{shinnaga04} Shinnaga, H.,
Moran, J.~M., Young, K.~H., \& Ho, P.~T.~P.\ 2004, \apjl, in press
(astro-ph/0408298)
\bibitem[Shinnaga et al.(2003)]{shinnaga03} Shinnaga, H., 
Claussen, M.~J., Lim, J., Dinh-van-Trung, \& Tsuboi, M.\ 2003, ASSL 
Vol.~283: Mass-Losing Pulsating Stars and their Circumstellar Matter, 393 
\bibitem[Smith et al.(2001)]{smith01} Smith, N., Humphreys, 
R.~M., Davidson, K., Gehrz, R.~D., Schuster, M.~T., \& Krautter, J.\ 2001, 
\aj, 121, 1111
\bibitem[Sopka et al.(1985)]{sopka85} Sopka, R.~J., Hildebrand, 
R., Jaffe, D.~T., Gatley, I., Roellig, T., Werner, M., Jura, M., \& 
Zuckerman, B.\ 1985, \apj, 294, 242
\bibitem[Strom, Johnston, Verbunt, \& Aschenbach(1995)]{strom95} 
Strom, R., Johnston, H.~M., Verbunt, F., \& Aschenbach, B.\ 1995, 
\nat, 373, 590 
\bibitem[Tokunaga(1984)]{tokunaga84} Tokunaga, A.~T.\ 1984, \aj, 
89, 172 
\bibitem[Tsuji(1964)]{tsuji64} Tsuji, T.\ 1964, Annals of the 
Tokyo Astronomical Observatory, 9
\bibitem[Wallerstein(1978)]{wallerstein78} Wallerstein, G.\ 1978, The 
Observatory, 98, 224 
\bibitem[Wang, Howell, H{\" o}flich, \& Wheeler(2001)]{wang01} 
Wang, L., Howell, D.~A., H{\" o}flich, P., \& Wheeler, J.~C.\ 2001, \apj, 
550, 1030 

\end{thebibliography}
\end{document}